\documentclass[10pt,a4paper,onecolumn]{article}
\usepackage{marginnote}
\usepackage{graphicx}
\usepackage{xcolor}
\usepackage{authblk,etoolbox}
\usepackage{titlesec}
\usepackage{calc}
\usepackage{tikz}
\usepackage{hyperref}
\hypersetup{colorlinks,breaklinks,
            urlcolor=[rgb]{0.0, 0.5, 1.0},
            linkcolor=[rgb]{0.0, 0.5, 1.0}}
\usepackage{caption}
\usepackage{tcolorbox}
\usepackage{amssymb,amsmath}
\usepackage{ifxetex,ifluatex}
\usepackage{seqsplit}
\usepackage{xstring}

\usepackage{float}
\let\origfigure\figure
\let\endorigfigure\endfigure
\renewenvironment{figure}[1][2] {
    \expandafter\origfigure\expandafter[H]
} {
    \endorigfigure
}

\usepackage{fixltx2e} 
\usepackage[
  backend=biber,
]{biblatex}
\bibliography{s4cmb\_paper.bib}

\let\textttOrig=\texttt
\def\texttt#1{\expandafter\textttOrig{\seqsplit{#1}}}
\renewcommand{\seqinsert}{\ifmmode
  \allowbreak
  \else\penalty6000\hspace{0pt plus 0.02em}\fi}

\makeatletter
\let\href@Orig=\href
\def\href@Urllike#1#2{\href@Orig{#1}{\begingroup
    \def\Url@String{#2}\Url@FormatString
    \endgroup}}
\def\href@Notdoi#1#2{\def\tempa{#1}\def\tempb{#2}%
  \ifx\tempa\tempb\relax\href@Urllike{#1}{#2}\else
  \href@Orig{#1}{#2}\fi}
\def\href#1#2{%
  \IfBeginWith{#1}{https://doi.org}%
  {\href@Urllike{#1}{#2}}{\href@Notdoi{#1}{#2}}}
\makeatother

\usepackage[top=3.5cm, bottom=3cm, right=1.5cm, left=1.0cm,
            headheight=2.2cm, reversemp, includemp, marginparwidth=4.5cm]{geometry}

\titleformat{\section}
  {\normalfont\sffamily\Large\bfseries}
  {}{0pt}{}
\titleformat{\subsection}
  {\normalfont\sffamily\large\bfseries}
  {}{0pt}{}
\titleformat{\subsubsection}
  {\normalfont\sffamily\bfseries}
  {}{0pt}{}
\titleformat*{\paragraph}
  {\sffamily\normalsize}

\usepackage{fancyhdr}
\makeatletter
\let\ps@plain\ps@fancy
\fancyheadoffset[L]{4.5cm}
\fancyfootoffset[L]{4.5cm}

\definecolor{linky}{rgb}{0.0, 0.5, 1.0}

\newtcolorbox{repobox}
   {colback=red, colframe=red!75!black,
     boxrule=0.5pt, arc=2pt, left=6pt, right=6pt, top=3pt, bottom=3pt}

\newcommand{\ExternalLink}{%
   \tikz[x=1.2ex, y=1.2ex, baseline=-0.05ex]{%
       \begin{scope}[x=1ex, y=1ex]
           \clip (-0.1,-0.1)
               --++ (-0, 1.2)
               --++ (0.6, 0)
               --++ (0, -0.6)
               --++ (0.6, 0)
               --++ (0, -1);
           \path[draw,
               line width = 0.5,
               rounded corners=0.5]
               (0,0) rectangle (1,1);
       \end{scope}
       \path[draw, line width = 0.5] (0.5, 0.5)
           -- (1, 1);
       \path[draw, line width = 0.5] (0.6, 1)
           -- (1, 1) -- (1, 0.6);
       }
   }

\patchcmd{\@maketitle}{center}{flushleft}{}{}
\patchcmd{\@maketitle}{center}{flushleft}{}{}
\patchcmd{\@maketitle}{\LARGE}{\LARGE\sffamily}{}{}
\def\maketitle{{%
  
  \AB@maketitle}}
\makeatletter
\renewcommand\AB@affilsepx{ \protect\Affilfont}
\renewcommand\AB@affilnote[1]{{\bfseries #1}\hspace{3pt}}
\renewcommand{\affil}[2][]%
   {\newaffiltrue\let\AB@blk@and\AB@pand
      \if\relax#1\relax\def\AB@note{\AB@thenote}\else\def\AB@note{#1}%
        \setcounter{Maxaffil}{0}\fi
        \begingroup
        \let\href=\href@Orig
        \let\texttt=\textttOrig
        \let\protect\@unexpandable@protect
        \def\thanks{\protect\thanks}\def\footnote{\protect\footnote}%
        \@temptokena=\expandafter{\AB@authors}%
        {\def\\{\protect\\\protect\Affilfont}\xdef\AB@temp{#2}}%
         \xdef\AB@authors{\the\@temptokena\AB@las\AB@au@str
         \protect\\[\affilsep]\protect\Affilfont\AB@temp}%
         \gdef\AB@las{}\gdef\AB@au@str{}%
        {\def\\{, \ignorespaces}\xdef\AB@temp{#2}}%
        \@temptokena=\expandafter{\AB@affillist}%
        \xdef\AB@affillist{\the\@temptokena \AB@affilsep
          \AB@affilnote{\AB@note}\protect\Affilfont\AB@temp}%
      \endgroup
       \let\AB@affilsep\AB@affilsepx
}
\makeatother

\renewcommand\Affilfont{\sffamily\small\mdseries}
\ifnum 0\ifxetex 1\fi\ifluatex 1\fi=0 
  \usepackage[T1]{fontenc}
  \usepackage[utf8]{inputenc}

\else 
  \ifxetex
    \usepackage{mathspec}
  \else
    \usepackage{fontspec}
  \fi
  \defaultfontfeatures{Ligatures=TeX,Scale=MatchLowercase}

\fi
\IfFileExists{upquote.sty}{\usepackage{upquote}}{}
\IfFileExists{microtype.sty}{%
\usepackage{microtype}
\UseMicrotypeSet[protrusion]{basicmath} 
}{}

\usepackage{hyperref}
\hypersetup{unicode=true,
            pdftitle={Simulating instrumental systematics of Cosmic Microwave Background experiments with s4cmb},
            pdfborder={0 0 0},
            breaklinks=true}
\let\addcontentslineOrig=\addcontentsline
\def\addcontentsline#1#2#3{\bgroup
  \let\texttt=\textttOrig\addcontentslineOrig{#1}{#2}{#3}\egroup}
\let\markbothOrig\markboth
\def\markboth#1#2{\bgroup
  \let\texttt=\textttOrig\markbothOrig{#1}{#2}\egroup}
\let\markrightOrig\markright
\def\markright#1{\bgroup
  \let\texttt=\textttOrig\markrightOrig{#1}\egroup}

\usepackage{graphicx,grffile}
\makeatletter
\def\maxwidth{\ifdim\Gin@nat@width>\linewidth\linewidth\else\Gin@nat@width\fi}
\def\maxheight{\ifdim\Gin@nat@height>\textheight\textheight\else\Gin@nat@height\fi}
\makeatother
\setkeys{Gin}{width=\maxwidth,height=\maxheight,keepaspectratio}
\IfFileExists{parskip.sty}{%
\usepackage{parskip}
}{
\setlength{\parindent}{0pt}
\setlength{\parskip}{6pt plus 2pt minus 1pt}
}
\providecommand{\tightlist}{%
  \setlength{\itemsep}{0pt}\setlength{\parskip}{0pt}}
\ifx\paragraph\undefined\else
\let\oldparagraph\paragraph
\renewcommand{\paragraph}[1]{\oldparagraph{#1}\mbox{}}
\fi
\ifx\subparagraph\undefined\else
\let\oldsubparagraph\subparagraph
\renewcommand{\subparagraph}[1]{\oldsubparagraph{#1}\mbox{}}
\fi

\title{Simulating instrumental systematics of Cosmic Microwave Background
experiments with s4cmb}

        \author[1, 3]{Giulio Fabbian\footnote{Corresponding author.}}
          \author[2]{Julien Peloton\footnote{Corresponding author.}}
    
      \affil[1]{Department of Physics \& Astronomy, University of Sussex, Brighton BN1
9QH, UK}
      \affil[2]{Université Paris-Saclay, CNRS/IN2P3, IJCLab, Orsay, France}
      \affil[3]{School of Physics and Astronomy, Cardiff University, The Parade,
Cardiff, CF24 3AA, UK}
  \date{\vspace{-5ex}}

\begin{document}
\maketitle

\marginpar{
  \sffamily\small

  {\bfseries DOI:} \href{https://doi.org/10.21105/joss.03022}{\color{linky}{10.21105/joss.03022}}

  \vspace{2mm}

  {\bfseries Software}
  \begin{itemize}
    \setlength\itemsep{0em}
    \item \href{https://github.com/openjournals/joss-reviews/issues/3022}{\color{linky}{Review}} \ExternalLink
    \item \href{https://github.com/JulienPeloton/s4cmb}{\color{linky}{Repository}} \ExternalLink
    \item \href{https://doi.org/10.5281/zenodo.4659617}{\color{linky}{Archive}} \ExternalLink
  \end{itemize}

  \vspace{2mm}
 
  {\bfseries Editor:} \href{http://stanford.edu/~mbobra/}{\color{linky}{Monica Bobra}} \ExternalLink
  
  \vspace{2mm}

  {\bfseries Reviewers}
  \begin{itemize}
    \setlength\itemsep{0em}
    \item \href{https://github.com/Christovis}{\color{linky}{@Christovis}} \ExternalLink
    \item \href{https://github.com/changhoonhahn}{\color{linky}{@changhoonhahn}} \ExternalLink
  \end{itemize}
  
  \vspace{2mm}  

  {\bfseries Submitted:} 02 February 2021\\
  {\bfseries Published:} 06 April 2021

  \vspace{2mm}
  {\bfseries License}\\
  Authors of papers retain copyright and release the work under a Creative Commons Attribution 4.0 International License (\href{https://creativecommons.org/licenses/by/4.0/}{\color{linky}{CC BY 4.0}}).
}

\hypertarget{summary}{%
\section{Summary}\label{summary}}

The observation of cosmic microwave background (CMB) anisotropies is one
of the key probes of the standard cosmological model (\hyperlink{ref-hu-dodelson2002}{Hu and Dodelson
2002}). The weak CMB polarization signal in particular can provide a new
window to study the process of the growth of structures in the universe
(galaxies, galaxy clusters, etc.) through weak gravitational lensing as
well as on the physics of the inflationary epoch in the primordial
universe. The low amplitude of CMB polarization has pushed CMB science
toward the construction of increasingly sensitive experiments observing
in multiple frequencies and employing telescopes with complex optical
designs and focal planes with thousands of bolometric detectors
operating in cryogenic environments (\hyperlink{ref-Staggs_2018}{Staggs, Dunkley, and Page 2018}). To
fully utilize the sensitivity of these instruments for cosmology, the
instrumental systematic effects must be well-characterized, understood,
and mitigated in the instrument design and through the analysis of the
acquired data.

\hypertarget{statement-of-need}{%
\section{Statement of need}\label{statement-of-need}}

The \texttt{s4cmb} (systematics for CMB) package is a Python package
developed to study the impact of instrumental systematic effects on
measurements of CMB experiments based on bolometric detector technology.
\texttt{s4cmb} provides a unified framework to simulate raw data streams
in the time domain (TODs) acquired by CMB experiments scanning the sky,
and to inject in these realistic instrumental systematics effect. The
development of \texttt{s4cmb} is built on experience and needs of the
analysis of data of the Polarbear ground-based experiment (\hyperlink{ref-pb2014}{Polarbear
Collaboration et al. 2014}, \hyperlink{ref-Ade:2017uvt}{2017}). It is designed to analyze real data,
to guide the design of future instruments that require the estimation of
specific systematic effects, as well as to increase the realism of
simulated data sets required in the development of data analysis
methods. The package has already been used in a number of scientific
(\hyperlink{ref-Puglisi:2018txk}{Puglisi et al. 2018}; \hyperlink{ref-mirmelstein2020}{Mirmelstein et al. 2020}) and technical
publications (\hyperlink{ref-Salatino:2018voz}{Salatino et al. 2018}; \hyperlink{ref-Crowley:2018eib}{Crowley et al. 2018}; \hyperlink{ref-Gallardo:2018rix}{Gallardo et al.
2018}; \hyperlink{ref-Bryan:2018mva}{Bryan et al. 2018}). It adopts several commonly used libraries in
astronomy (\texttt{astropy} (\hyperlink{ref-astropy}{Astropy Collaboration 2013}),
\texttt{healpy} (\hyperlink{ref-healpy}{Zonca et al. 2019}), \texttt{ephem} (\hyperlink{ref-pyephem}{Rhodes, n.d.}),
\texttt{pyslalib} (\hyperlink{ref-pyslalib}{Ransom 2010})) and uses functions based on low-level
languages wrapped in Python (e.g., Fortran with \texttt{f2py}) for
speeding up the most critical part of the code without losing the
flexibility provided by a simple Python user-friendly interface.
\texttt{s4cmb} is designed to be employed on systems of varying scale,
from laptops to parallel supercomputing platforms, thanks to its
internal Message Passing Interface (MPI) support (\hyperlink{ref-10.1145ux2f169627.169855}{The MPI Forum 1993}).
We also support packaging the entire application into a Docker container
for portability. The simplicity of the \texttt{s4cmb} framework allows a
user to easily add new instrumental systematics to be simulated
according to the user's needs. As far as we know, s4cmb is the only
dedicated package that enables the study of a wide range of instrumental
simulations, from the instrument to the sky map, while being publicly
available. For more general purposes, including some instrumental
systematic effect simulations, users might also consider the use of
\texttt{TOAST} (\hyperlink{ref-theodore_kisner_2020_4270476}{Kisner et al. 2020}), a software framework to simulate
and process timestream data collected by telescopes focusing on
efficient TOD manipulation on massively parallel architectures.

\hypertarget{package-structure-and-functionalities}{%
\section{Package structure and
functionalities}\label{package-structure-and-functionalities}}

One of the key feature of \texttt{s4cmb} is to be able to make robust
simulations while being fast to run and easy to use. Simulations using
\({\cal{O}}(10^3)\) detectors observing the sky for 5 hours with a data
sampling rate of \(100\)Hz and reconstructing a sky map can be run on a
single core in less than 10 minutes on a laptop. The modules
implementing the major functional blocks of the library are (see
\autoref{fig:s4cmb}):

\begin{figure}
\centering
\includegraphics[width=1\textwidth,height=\textheight]{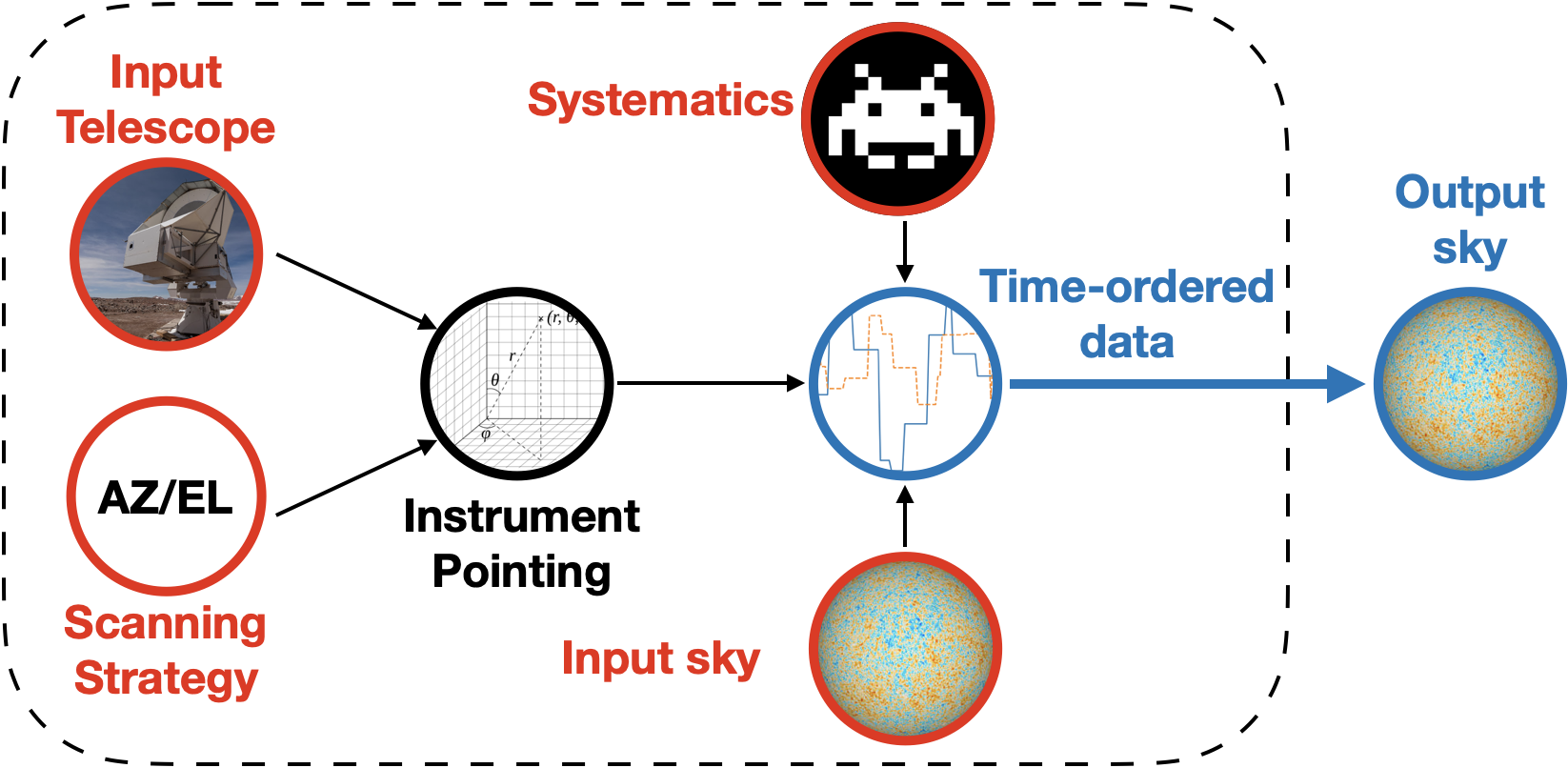}
\caption{Schematic structure of \texttt{s4cmb}. Objects defined by the
user are marked in red. From these, TODs are generated for the duration
of each observation described by the scanning strategy. These can then
be modified by introducing instrumental effects. The output of the code
(blue) are the perturbed TODs or sky maps reconstructed from TODs to be
employed in subsequent analysis steps (e.g., the computation of their
angular power spectrum).\label{fig:s4cmb}}
\end{figure}

\begin{itemize}
\tightlist
\item
  \texttt{instrument.py} contains the class describing the CMB
  instrument in terms of position of its detectors in the focal plane,
  their optical beam shape, and wiring in the readout electronics. It
  supports the most common focal plane designs employing multifrequency
  detectors and polarization modulation hardware such as stepped or
  continuously rotating half-wave plates (HWPs).
\item
  \texttt{scanning\_strategy.py} describes the schedule of the
  instrument observations and the motion of the instrument in terms of
  azimuth and elevation position at the telescope location on Earth as a
  function of time. The schedule is divided in minimal units (scans)
  during which a telescope motion is repeated for its given duration
  (e.g., a back and forth motion for a fixed distance in azimuth at a
  constant elevation.) The code parallelisation is done over scans, so
  that proportionally increasing the number of scans and the number of
  MPI workers keeps the runtime constant.
\item
  \texttt{input\_sky.py} contains the class describing the input sky
  signal model in HEALPix pixelization (\hyperlink{ref-healpix}{Górski et al. 2005}). This can
  include multiple components (CMB, Galactic emissions, etc.) and can be
  read from an external file or be synthesized on the fly from its
  harmonic coefficients. The input sky needs be convolved with the main
  optical beam of the instrument \(B\) so that the observed sky Stokes
  parameters \(X^{\rm obs}\in\{I, Q, U\}\) are related to their true
  value on sky a \(X^{\rm obs} \equiv B \circledast X\).
\item
  \texttt{tod.py} includes the class to generate and handle TODs from an
  input sky signal map, an instrument design, and a scanning strategy,
  together with basic tools to reconstruct a sky map from TODs in order
  to self consistently mimic the basic data analysis pipeline of CMB
  experiments.
\end{itemize}

\begin{figure}
\centering
\includegraphics{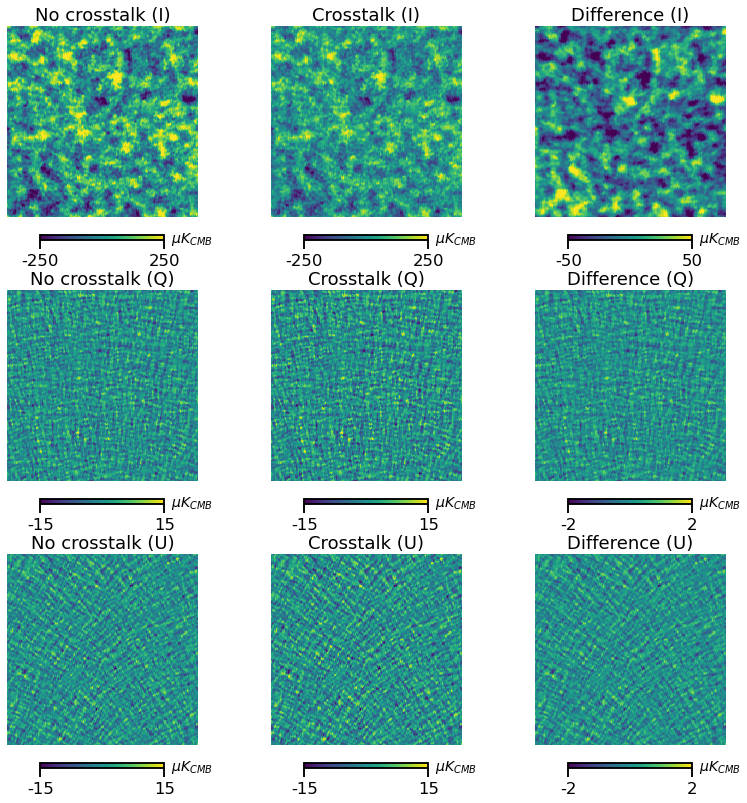}
\caption{Impact of electrical crosstalk in the detectors' readout
electronics in CMB observations. For each Stokes parameter (top to
bottom row) we show the input sky map (left), the sky map reconstructed
from TOD affected by crosstalk (middle), and their difference
(right).\label{fig:crosstalk}}
\end{figure}

In the absence of instrumental systematic effects, the TOD of a single
detector \(d\) is modeled as:

\begin{equation}
d_t  = g_t[I^{\rm obs}(\hat{\mathbf{n}}_t)+\cos2\psi_t Q^{\rm obs}(\hat{\mathbf{n}}_t) + \sin2\psi_t U^{\rm obs}(\hat{\mathbf{n}}_t)] +n_t,
\label{eq:datamodel}
\end{equation}

where a \(t\) subscript denotes a time-dependent quantity.
\(\mathbf{n}\) is the vector describing the detector pointing and
\(\psi\) is the polarization angle that describes the effective
orientation of a polarization sensitive detector with respect to the
input sky coordinate system. This depends on the orientation of the
detector itself but also on the orientation of optical elements that
modulate the incoming polarized signal (e.g., HWPs). In absence of
systematics, the calibration factor (gain) is \(g_t=1\) to preserve the
calibration of the input signal. The exact value of the pixelized input
Stokes parameters at \(\mathbf{n}_t\) is determined through nearest grid
point interpolation. White or correlated noise \(n_t\) can be added to
TOD according to the instrument specifications.

\hypertarget{instrumental-systematic-effects}{%
\subsubsection{Instrumental systematic
effects}\label{instrumental-systematic-effects}}

The instrumental systematic effects implemented in \texttt{s4cmb} are
available in the \texttt{systematics.py} module. This includes
electrical crosstalk in the detectors' readout electronics, gain
misestimation or drifting of their values in time, pointing errors,
distortions of the beam shape compared to the shape of the expected beam
\(B\) used to convolve the input sky and misestimation of their position
in the focal plane, errors in the polarization angle estimation. These
effectively modify on the fly the \(\psi_t\), \(\mathbf{n}_t\), \(g_t\)
in \autoref{eq:datamodel} compared to their expected value determined by
the instrument design and scanning strategy when creating the TODs.
Effects of beam distortions are conversely modeled through a Taylor
expansion of \(B\). Further details of the mathematical modeling of
systematics are given in \hyperlink{ref-mirmelstein2020}{Mirmelstein et al. (2020)} and \hyperlink{ref-bicep-beam}{Bicep2
Collaboration et al. (2015)}. We encourage users to add more effects and
integrate their work in the package through pull requests on GitHub.

\hypertarget{bootcamp}{%
\section{Bootcamp}\label{bootcamp}}

We release a
\href{https://github.com/JulienPeloton/s4cmb-resources}{bootcamp}
dedicated to the package in two parts (beginners and advanced users)
that include notebooks describing the basic parts of the API, and
providing ready-to-use examples for the major code functionalities. A
notebook to create \autoref{fig:crosstalk} can be found
\href{https://github.com/JulienPeloton/s4cmb-resources/blob/master/Part1/s4cmb_crosstalk_05-joss.ipynb}{here}.

\hypertarget{acknowledgements}{%
\section{Acknowledgements}\label{acknowledgements}}

We thank Neil Goeckner Wald for contributing to a large part of the
scanning strategy module, and the Polarbear collaboration for fostering
the developments that led to this package. We acknowledge support from
the European Research Council under the European Union's Seventh
Framework Programme (FP/2007-2013) / ERC Grant Agreement No.
{[}616170{]}. GF acknowledges the support of the European Research
Council under the Marie Skłodowska Curie actions through the Individual
Global Fellowship No.~892401 PiCOGAMBAS.

\hypertarget{references}{%
\section*{References}\label{references}}
\addcontentsline{toc}{section}{References}

\hypertarget{refs}{}
\leavevmode\hypertarget{ref-astropy}{}%
Astropy Collaboration. 2013. ``Astropy: A community Python package for
astronomy.'' \emph{Astron. Astrophys.} 558 (October).
\url{https://doi.org/10.1051/0004-6361/201322068}.

\leavevmode\hypertarget{ref-bicep-beam}{}%
Bicep2 Collaboration, P. A. R. Ade, R. W. Aikin, D. Barkats, S. J.
Benton, C. A. Bischoff, J. J. Bock, et al. 2015. ``Bicep2 III:
Instrumental Systematics.'' \emph{Astrophys. J.} 814 (2): 110.
\url{https://doi.org/10.1088/0004-637X/814/2/110}.

\leavevmode\hypertarget{ref-Bryan:2018mva}{}%
Bryan, Sean A., Sara M. Simon, Martina Gerbino, Grant Teply, Amir Ali,
Yuji Chinone, Kevin Crowley, et al. 2018. ``Development of calibration
strategies for the Simons Observatory.'' In \emph{Proc. SPIE Int. Soc.
Opt. Eng.}, edited by George Z. Angeli and Philippe Dierickx,
10708:1070840. \url{https://doi.org/10.1117/12.2313832}.

\leavevmode\hypertarget{ref-Crowley:2018eib}{}%
Crowley, Kevin T., Sara M. Simon, Max Silva-Feaver, Neil Goeckner-Wald,
Aamir Ali, Jason Austermann, Michael L. Brown, et al. 2018. ``Studies of
systematic uncertainties for Simons Observatory: detector array
effects.'' \emph{Proc. SPIE Int. Soc. Opt. Eng.} 10708: 107083Z.
\url{https://doi.org/10.1117/12.2313414}.

\leavevmode\hypertarget{ref-Gallardo:2018rix}{}%
Gallardo, Patricio A., Jon Gudmundsson, Brian J. Koopman, Frederick T.
Matsuda, Sara M. Simon, Aamir Ali, Sean Bryan, et al. 2018. ``Studies of
Systematic Uncertainties for Simons Observatory: Optical Effects and
Sensitivity Considerations.'' Edited by George Z. Angeli and Philippe
Dierickx. \emph{Proc. SPIE Int. Soc. Opt. Eng.} 10708: 107083Y.
\url{https://doi.org/10.1117/12.2312971}.

\leavevmode\hypertarget{ref-healpix}{}%
Górski, K. M., E. Hivon, A. J. Banday, B. D. Wandelt, F. K. Hansen, M.
Reinecke, and M. Bartelmann. 2005. ``HEALPix: A Framework for
High-Resolution Discretization and Fast Analysis of Data Distributed on
the Sphere.'' \emph{Astrophys. J.} 622 (2): 759--71.
\url{https://doi.org/10.1086/427976}.

\leavevmode\hypertarget{ref-hu-dodelson2002}{}%
Hu, Wayne, and Scott Dodelson. 2002. ``Cosmic Microwave Background
Anisotropies.'' \emph{Ann. Rev Astron. Astrophys.} 40 (January):
171--216. \url{https://doi.org/10.1146/annurev.astro.40.060401.093926}.

\leavevmode\hypertarget{ref-theodore_kisner_2020_4270476}{}%
Kisner, Theodore, Reijo Keskitalo, Andrea Zonca, Giuseppe Puglisi, and
Julian Borrill. 2020. \emph{Hpc4cmb/Toast: Small Fixes Discovered Post
2.3.11} (version 2.3.12). Zenodo.
\url{https://doi.org/10.5281/zenodo.4270476}.

\leavevmode\hypertarget{ref-mirmelstein2020}{}%
Mirmelstein, Mark, Giulio Fabbian, Antony Lewis, and Julien Peloton.
2020. ``Instrumental systematics biases in CMB lensing reconstruction: a
simulation-based assessment,'' November, \href{https://arxiv.org/abs/2011.13910}{\texttt{arXiv:2011.13910}}.

\leavevmode\hypertarget{ref-Ade:2017uvt}{}%
Polarbear Collaboration, P. A. R. Ade, M. Aguilar, Y. Akiba, K. Arnold,
C. Baccigalupi, D. Barron, et al. 2017. ``A Measurement of the Cosmic
Microwave Background \(B\)-Mode Polarization Power Spectrum at
Sub-Degree Scales from 2 years of Polarbear Data.'' \emph{Astrophys. J.}
848 (2): 121. \url{https://doi.org/10.3847/1538-4357/aa8e9f}.

\leavevmode\hypertarget{ref-pb2014}{}%
Polarbear Collaboration, P. A. R. Ade, Y. Akiba, A. E. Anthony, K.
Arnold, M. Atlas, D. Barron, et al. 2014. ``A Measurement of the Cosmic
Microwave Background B-Mode Polarization Power Spectrum at Sub-Degree
Scales with Polarbear.'' \emph{Astrophys. J.} 794 (2): 171.
\url{https://doi.org/10.1088/0004-637X/794/2/171}.

\leavevmode\hypertarget{ref-Puglisi:2018txk}{}%
Puglisi, Giuseppe, Davide Poletti, Giulio Fabbian, Carlo Baccigalupi,
Luca Heltai, and Radek Stompor. 2018. ``Iterative map-making with
two-level preconditioning for polarized cosmic microwave background data
sets: A worked example for ground-based experiments.'' \emph{Astron.
Astrophys.} 618: A62. \url{https://doi.org/10.1051/0004-6361/201832710}.

\leavevmode\hypertarget{ref-pyslalib}{}%
Ransom, S. M. 2010. ``PySLALIB.'' \emph{GitHub Repository}. GitHub.
\url{https://github.com/scottransom/pyslalib}.

\leavevmode\hypertarget{ref-pyephem}{}%
Rhodes, B. n.d. ``PyEphem.'' \emph{GitHub Repository}. GitHub.
\url{https://github.com/brandon-rhodes/pyephem}.

\leavevmode\hypertarget{ref-Salatino:2018voz}{}%
Salatino, Maria, Jacob Lashner, Martina Gerbino, Sara M. Simon, Joy
Didier, Aamir Ali, Peter C. Ashton, et al. 2018. ``Studies of systematic
uncertainties for Simons Observatory: polarization modulator related
effects.'' Edited by Jonas Zmuidzinas and Jian-Rong Gao. \emph{Proc.
SPIE Int. Soc. Opt. Eng.} 10708 (July): 1070848.
\url{https://doi.org/10.1117/12.2312993}.

\leavevmode\hypertarget{ref-Staggs_2018}{}%
Staggs, Suzanne, Jo Dunkley, and Lyman Page. 2018. ``Recent Discoveries
from the Cosmic Microwave Background: A Review of Recent Progress.''
\emph{Reports on Progress in Physics} 81 (4). IOP Publishing: 044901.
\url{https://doi.org/10.1088/1361-6633/aa94d5}.

\leavevmode\hypertarget{ref-10.1145ux2f169627.169855}{}%
The MPI Forum, CORPORATE. 1993. ``MPI: A Message Passing Interface.'' In
\emph{Proceedings of the 1993 Acm/Ieee Conference on Supercomputing},
878--83. Supercomputing '93. New York, NY, USA: Association for
Computing Machinery. \url{https://doi.org/10.1145/169627.169855}.

\leavevmode\hypertarget{ref-healpy}{}%
Zonca, Andrea, Leo P. Singer, Daniel Lenz, Martin Reinecke, Cyrille
Rosset, Eric Hivon, and Krzysztof M. Gorski. 2019. ``Healpy: Equal Area
Pixelization and Spherical Harmonics Transforms for Data on the Sphere
in Python.'' \emph{Journal of Open Source Software} 4 (35). The Open
Journal: 1298. \url{https://doi.org/10.21105/joss.01298}.

\end{document}